\documentstyle[epsfig]{elsart}

\begin{document}

\vspace*{2cm}

\begin{center}
{\Large {   \bf
            Systematics of Inclusive Photon Production in
       158$\cdot A$ GeV Pb Induced Reactions on Ni, Nb, and Pb Targets
\\
}}
\end{center}

\bigskip

{\small{
\noindent
M.M.~Aggarwal,$^{1}$
A.~Agnihotri,$^{2}$
Z.~Ahammed,$^{3}$
A.L.S.~Angelis,$^{4}$ 
V.~Antonenko,$^{5}$ 
V.~Arefiev,$^{6}$
V.~Astakhov,$^{6}$
V.~Avdeitchikov,$^{6}$
T.C.~Awes,$^{7}$
P.V.K.S.~Baba,$^{8}$
% ...
}

\medskip

\begin{center}
{\large {(WA98 Collaboration)}}
\end{center}

\bigskip

\begin{frontmatter}

{\small{
{$^{1}$~University of Panjab, Chandigarh 160014, India}
{$^{2}$~University of Rajasthan, Jaipur 302004, Rajasthan,
  India}
{$^{3}$~Variable Energy Cyclotron Centre,  Calcutta 700064,
  India}
{$^{4}$~University of Geneva, CH-1211 Geneva 4,Switzerland}
{$^{5}$~RRC (Kurchatov Institute), RU-123182 Moscow, Russia}
{$^{6}$~Joint Institute for Nuclear Research, RU-141980 Dubna,
  Russia}
{$^{7}$~Oak Ridge National Laboratory, Oak Ridge, Tennessee
  37831-6372, USA}
{$^{8}$~University of Jammu, Jammu 180001, India}
{$^{9}$~University of M{\"u}nster, D-48149 M{\"u}nster,
  Germany}
{$^{10}$~SUBATECH, Ecole des Mines, Nantes, France}
{$^{11}$~Gesellschaft f{\"u}r Schwerionenforschung (GSI),
  D-64220 Darmstadt, Germany}
{$^{12}$~Universiteit Utrecht/NIKHEF, NL-3508 TA Utrecht, The
  Netherlands}
{$^{13}$~Lund University, SE-221 00 Lund, Sweden}
{$^{14}$~University of Tsukuba, Ibaraki 305, Japan}
{$^{15}$~Nuclear Physics Institute, CZ-250 68 Rez, Czech Rep.}
{$^{16}$~KVI, University of Groningen, NL-9747 AA Groningen,
  The Netherlands}
{$^{17}$~Institute for Nuclear Studies, 00-681 Warsaw, Poland}
{$^{18}$~MIT Cambridge, MA 02139, USA}
{$^{19}$~Institute of Physics, 751-005  Bhubaneswar, India}
{$^{20}$~University of Tennessee, Knoxville, Tennessee 37966,
  USA}
}}           

\begin{abstract} 
    The multiplicity of inclusive photons has
    been measured on an event-by-event basis 
    for 158$\cdot A$ GeV Pb induced reactions on Ni, Nb, and Pb targets.
    The systematics of the pseudorapidity 
    densities at midrapidity ($\rho_{max}$)
    and the width of the pseudorapidity distributions 
    have been studied for varying centralities for these collisions.
    A power law fit to the photon yield as a function of the
    number of participating nucleons gives a 
    value of $1.13\pm0.03$ for the exponent.
    The mean transverse momentum, 
    $\langle p_T\rangle$, 
    of photons determined from the ratio of the measured electromagnetic
    transverse energy and photon multiplicity, remains almost
    constant with increasing $\rho_{max}$.
    Results are compared with model predictions.
\end{abstract}
\end{frontmatter}

\normalsize

\section{Introduction}
     The primary goal of ultra-relativistic heavy ion experiments 
     is to study nuclear matter under extreme conditions, in which 
     hadronic matter is expected to undergo a phase transition
     to a new state of matter, the Quark-Gluon-Plasma (QGP).
     For a thermalized system undergoing a phase transition, the variation of
     the temperature with entropy density is interesting as the
     temperature is expected to increase while
     below the transition, remain constant during the transition,
     and then increase again \cite{shuryak,vanhove}.  
     Temperature fluctuations have also been proposed
     as a signature of the existence of a tricritical point in QCD
     \cite{step}.
     These behaviours can 
     be studied by two experimentally measured quantities, {\it viz.},
     the mean transverse momentum, $\langle p_T\rangle$, and the pseudorapidity
     density at midrapidity, $\rho_{max}$, for 
     varying impact parameter, or centrality, for a 
     number of colliding systems.
     These variables also provide 
     additional information to characterize the evolving system.
     $\rho_{max}$ provides a measure of the energy density which is
     important to understand the reaction dynamics \cite{bjorken,kataja}.
     In addition, 
     the change in shape of the pseudorapidity distribution should be
     investigated in detail because it may provide a clue to the formation
     of the QGP phase \cite{sarkar,dumitru}.

     Except for a few measurements 
     of photon \cite{wa93bgo,wa93pt,wa93eta,wa80_2}
     and neutral meson \cite{wa80_pi0,wa80_eta,wa80_pi02,wa98_pi0}
     distributions
     reported earlier, most  studies have been restricted to
     charged particle measurements (a review of charged particle
     measurements can be found in \cite{stachel,aggar}),
     due to the difficulty of precise measurements of photon distributions.
     The inclusive photons
     provide a picture of the system at freezeout since the 
     majority of the photons emitted from the reaction 
     are decay products of produced particles, $viz.$, $\pi_0$ and $\eta$,
     (only a small fraction are emitted directly during 
     the initial stage of the collision\cite{photons}). 
     The shape and width of the pseudorapidity
     distributions of photons may be different
     compared to those of the charged particles.
     Photon multiplicity measurements
     have also become increasingly important because of the recent
     interest in simultaneous measurements of the multiplicity 
     of photons 
     and charged particles in the search for
     Disoriented Chiral Condensates (DCC)\cite{dcc,wa98dcc}.
     The formation of a DCC is expected to give rise 
     to large fluctuations in the relative  
     number of emitted charged particles and photons, analogous to the
     Centauro and the anti-Centauro types of events observed in cosmic
     ray experiments \cite{centauro}. 
     Photon measurements can also be used to study flow\cite{wa93flow}
     and intermittency behavior of events accompanying a possible phase
     transition. 


     In this letter, we present the first measurement of the photon multiplicity and 
     pseudorapidity distributions, together with the $\langle
     p_T\rangle$ of photons
     produced in collisions of 158$\cdot A$ GeV
     Pb with Ni, Nb, and Pb targets, carried out 
     in the WA98 experiment\cite{wa98} at the CERN-SPS.

\section{Experimental Setup}
    In the WA98 experiment, the main emphasis has been on high
    precision, simultaneous detection of both hadrons and photons.
    The experimental setup consists of large acceptance hadron and photon 
    spectrometers, detectors for charged particle and photon multiplicity 
    measurements, and calorimeters for transverse and forward energy 
    measurements. The present analysis makes use of 
    the photon multiplicity detector (PMD), 
    the midrapidity calorimeter (MIRAC) and the zero degree calorimeter (ZDC).

    The PMD was placed at a distance of 21.5 meters from the target.
    It was a preshower detector consisting of 3 radiation length ($X_0$)
    thick lead converter plates in
    front of an array of square scintillator pads of four sizes, varying 
    from 15 mm $\times$ 15 mm to 25 mm $\times$ 25 mm, placed in 28 box modules.
    Each box module consisted
    of a matrix of 38 $\times$ 50 pads and was read out using one image 
    intensifier + CCD camera system. The scintillation light was transmitted 
    to the readout device by using a short wavelength shifting 
    fiber spliced with a long EMA (extra-mural absorber) coated clear fiber. 
    The total light amplification of the readout 
    system was $\sim$40000. Digitization 
    of the CCD pixel charge was done by a set of custom built 
    fastbus modules employing an 8 bit 20 MHz Flash ADC system. 
    Details of the 
    design and characteristics of the PMD may be found in ref. \cite{wa98nim}.
    The results presented here make use 
    of the data from the central 22 box modules covering the
    pseudorapidity range of  $2.9\le \eta\le 4.2$.

    The MIRAC \cite{awes} was placed behind the PMD at 24.7 meters 
    from the target. It
    consisted of 30 stacks, each divided vertically into 6 towers, each of 
    size 20 cm $\times$ 20 cm, and segmented longitudinally into
    electromagnetic (EM) and hadronic sections. The depth of an EM section 
    is 15.6$X_0$ (equivalent to 51\% of an interaction length)
    which ensures essentially complete containment of
    the electromagnetic energy, with 97.4\% and 91.0\% containment calculated
    for 1~GeV and 30~GeV photons, respectively. Hadrons also deposit a sizable
    fraction of their energy in the EM section.
    The MIRAC measures both the transverse electromagnetic ($E_T^{em}$) 
    and hadronic ($E_T^{had}$) energies in the
    interval $3.5\le\eta\le 5.5$ with  
    a resolution of 17.9\%/$\sqrt E$
    and 46.1\%/$\sqrt E$, respectively, where $E$ is expressed in GeV.
    The ZDC measures the total forward energy, $E_F$, with a resolution
    of 80\%/$\sqrt E + 1.5\%$, where $E$ is expressed in GeV. 

\section{Data analysis}
\subsection{Event Selection}
    The data were taken during the December 1996 Pb beam period
    at the CERN SPS. The thicknesses of the three targets were 
    250~$\mu$m, 254~$\mu$m, and 213~$\mu$m for Ni, Nb, and Pb, respectively.
    The fundamental ``beam'' trigger condition consisted of a signal in
    a gas \v{C}erenkov start counter located 3.5 meters upstream
    of the target and no coincident signal in a
    veto counter with a 3 mm circular hole located 2.7
    meters upstream from the target.
    A beam trigger was considered to be a minimum-bias interaction if
    the transverse energy sum in the full MIRAC acceptance
    exceeded a low threshold.

    Beam pile up, where a second beam trigger
    occurs at a time when the detectors are integrating their signals
    from the triggered event, was rejected by (a) using the
    timing information from the trigger detectors, and
    (b) requiring that the sum of energies in the ZDC and the MIRAC were 
    within $3\sigma$ from the average.
    Downstream interactions were also rejected by 
    requiring a coincident signal from the forward hemisphere 
    of the Plastic Ball detector which surrounded the target.
    To correct for other sources of background, 
    data were also taken with no target in place.
    The target-out contributions were negligible except for the most
    peripheral reactions.

    The CCD readout of the PMD 
    was cleared every 10~$\mu$s using a clear pulse of 1~$\mu$s
    width generated every 10~$\mu$s. This ensured that there was no
    substantial noise buildup on the CCD pixels between successive event
    triggers. A gate of 2~$\mu$s around the clear pulse was used
    to veto partially or fully cleared events.
    The clear clock operated asynchronously and was vetoed with
    a 5.6~ms wide pulse when a valid trigger occurred to allow
    complete readout of all pixels.
    A further check on possible pileup in the CCD cameras
    was made by using a 10~$\mu$s range TDC to measure the time difference
    between the arrival of the last clear clock and any valid event trigger.
    Events with multiple interactions within the  10~$\mu$s
    between clear pulses were rejected in the offline analysis.


    The centrality of the interaction is determined by the
    total transverse energy
    measured in the MIRAC. For the $\langle p_T\rangle$ analysis,  
    which used the MIRAC data directly, the
    centrality was determined instead by the forward energy, $E_F$, 
    measured in the ZDC. 
    The centralities are 
    expressed as fractions of the minimum bias cross section 
    as a function of
    the measured total transverse energy or measured $E_F$. The most central
    selection corresponds to the top 5\% of the minimum bias cross section
    and the peripheral selection corresponds to the 
    lower 50$-$80\% range. Extreme
    peripheral events in the 80$-$100\% range were not analyzed.

\subsection{Data Reduction in the PMD}
    The digitized pixel charges 
    are processed by using a pixel-to-fiber map to form fiber signals 
    corresponding to
    each scintillator pad. The signals from several neighbouring 
    scintillator pads are combined to 
    form clusters, characterized by the total ADC content and the hit 
    positions. 
    On average, there is a 92\% probability for photons
    to shower in the lead converter and produce large
    signals. Compared to this, hadrons give a signal mostly
    corresponding to minimum ionizing particles (MIP). 
    The majority of the hadrons are rejected by applying a suitable threshold
    on the cluster signal. 
    A fraction of hadrons undergoing interaction in the lead converter
    produce signals larger than the threshold and appear as contaminants in
    the photon sample.  The number of clusters
    remaining above the hadron rejection threshold is termed as 
    $\gamma$-$like$.
    The characteristics of the preshower PMD are described by the following
    two quantities \cite{wa98nim}:
\begin{equation}
       \epsilon_\gamma  =  N^{\gamma,th} _{cls} / N_\gamma ^{inc}
\end{equation}
\begin{equation}
       f_p              =  N^{\gamma,th} _{cls} / N_{\gamma-like}
\end{equation}
     where $\epsilon_\gamma$ is the photon counting
     efficiency and $f_p$ is the fractional purity of the photon sample.
     \noindent $N_\gamma ^{inc}$ is the number of incident photons on the PMD
     and $N^{\gamma,th}_{cls}$ is the number of photon clusters above the 
     threshold. 
     Both $\epsilon_\gamma$ and $f_p$ are determined by a detailed 
     Monte-Carlo simulation
     using the VENUS~4.12 \cite{venus} 
     event generator with default parameter settings and the detector 
     simulation package, GEANT~3.21 \cite{geant}.
     The details of the simulation can be found in Ref. \cite{wa98nim}.
     No lower threshold on the energy spectrum of photons is applied in the
     simulation. The photon counting efficiency decreases
     with increasing hadron rejection threshold. 
     The purity improves significantly with
     increasing threshold only up to $\sim$~3 MIPs and then
     rather slowly at higher thresholds. For practical purposes a 3 MIP
     threshold appears as an optimum choice for hadron rejection. With this
     value the photon counting efficiencies for the central and peripheral cases are
     68\% and 73\%, respectively. The purity of the photon sample in the
     two cases are 65\% and 54\%, respectively.

     From the experimental data one determines $N_{\gamma-like}$, the
     number of clusters above the hadron rejection threshold. Using the
     estimated values of $\epsilon_\gamma$ and $f_p$, one obtains the
     number of photons incident on the detector in the event from 
     the relation:
\begin{equation}
       N_\gamma = N_{\gamma-like} \cdot f_p / \epsilon_\gamma
\end{equation}

\subsection{Extraction of $\langle p_T\rangle$ of photons}
    In a given event, the average transverse momentum of produced photons
    may be expressed as
\begin{equation}
         \langle p_T\rangle = \frac{E^{em}_T}{N_\gamma}
\end{equation}
    where $E^{em}_T$ is the transverse component of the electromagnetic energy,
    and $N_\gamma$ is the number of photons in a given $\eta$-region.
    In the WA98 experiment, 
    $E^{em}_T$ and $N_\gamma$ are measured with the MIRAC and the PMD,
    respectively, on an event-by-event
    basis.
    These detectors have complete overlap in azimuth in the region
    $3.5 \le \eta \le 4.0$. Hence the data in this region
    are used for computing the $\langle p_T\rangle$ using the above equation.

    In order to obtain the final $E^{em}_T$ for equation~(4), the measured
    electromagnetic energy in the MIRAC towers must be corrected for
    (1) the hadronic contribution to the EM section of the MIRAC, 
    and (2) the energy deposited in the lead converter of the PMD 
    because of its
    position in front of the MIRAC.  The final expression may be written as:
\begin{equation}
     E_T^{em} = \frac{\sum_{i=1}^{N} [E_i^{em}  - 
             f_h \cdot f_{bal} \cdot  \{E_i^{had}/(1-f_h)\}] 
\sin \theta_i}{1-f_{PMD}}
\end{equation}
where,
   $E_i^{em}$ and $E_i^{had}$ are the energies measured in the electromagnetic
   and hadronic sections of the MIRAC towers, 
   $f_h$ is the fraction of the hadronic energy deposited in the EM section,
   $f_{bal}$ is the balance factor taking into account the different responses
   for electromagnetic and hadronic particles in the EM section~\cite{awes},
   $N$ is the  number of  towers in the MIRAC within the given $\eta$ range,
   $\theta_i$ is the polar angle of the $i^{th}$ tower, and 
   $f_{PMD}$ is the fraction of the electromagnetic energy deposited in the
   lead converter of the PMD. The value of $f_{PMD}$ is found to be 15\%.
   Details of the corrections are similar to those of Ref. \cite{wa93pt}.

\subsection{Systematic Errors}

    Several different sources contribute to the final systematic error
    in the determination of the number of photons, $N_\gamma$.  
    The error due to the effect of clustering of the pad signals
    is the dominant one. 
    This error is determined from the simulation by comparing the
    number of known tracks on the PMD with the total number of 
    clusters obtained after clustering. 
    Apart from the effect of multiplicity as discussed in \cite{wa93eta},
    the arrangement of box modules in the present setup leads
    to splitting of clusters at the box boundaries.
    The net result is that
    the number of clusters exceeds the number of tracks with a deviation of
    3\% in the case of
    peripheral events to 7\% for high multiplicity central events.
    

    The number of $\gamma$-like clusters depends on the ADC value of
    the MIP peak as determined from Pb+Pb data.
    It has been estimated~\cite{wa98nim} that because of 
    an admixture of $\sim$20\%  photons in the MIP sample in the data,
    the extracted MIP ADC value is higher by
    2 ADC channels. This causes the extracted photon multiplicity
    to be lower by 2.5\%. We have included this as a source of
    systematic error.
    The error on $\epsilon_\gamma$ because of the variation in
    pad-to-pad gains is found to be less than 1\%.

    The purity factor, $f_p$, depends on the ratio of the number of photons
    and charged particles within the PMD coverage. The
    systematic error associated with this ratio has been
    studied by using the FRITIOF \cite{fritiof} event generator in addition
    to VENUS. The average photon multiplicity
    by using FRITIOF is found to be higher
    by about 4\% in peripheral and by 1\% in central collisions compared to
    using VENUS.

    The combined systematic error on the final 
    photon multiplicity is asymmetric and varies 
    with centrality of the reaction.
    The total systematic errors are $-$3.4\% and $+$7.5\% 
    for peripheral collisions
    and $-$7.1\% and $+$3.4\% for central collisions, varying little throughout
    the PMD acceptance. The negative error implies overestimation of
    number of photons.

    The photon counting efficiency determined in the present case relies on
    the energy spectra of photons as given by the VENUS event generator. As
    the conversion probability for low energy photons falls sharply 
    \cite{wa93nim} with decreasing energy below 500 MeV, the estimate of
    $\epsilon_\gamma$ 
    may be affected if the energy spectra in the actual case is
    different. 
    Preliminary measurements of the photon energy spectra with the
    WA98 lead glass spectrometer indicate
    that there is an
    enhancement of photons below $p_T=250$~MeV/c over that given by VENUS. 
    Taking into account this excess of low energy photons in the 
    PMD acceptance,
    the photon counting efficiency would be overestimated by
    2$-$9\% for central events
    and 3$-$13\% for peripheral events, the smaller value being for
    large pseudorapidity and the larger value being for the smaller
    pseudorapidity region of the PMD acceptance. The effect would
    be to increase the quoted PMD photon multiplicities. However, 
    this uncertainty in the photon counting efficiency due to the
    uncertainty in the photon spectrum has not been 
    included in the final errors. The WA98 low
    energy photon measurements will be described in a separate publication.


    The systematic error on the determination of $\langle p_T\rangle$ depends
    on the error in both $E_T^{em}$ and $N_\gamma$. 
    The major source of errors
    in $E_T^{em}$ are the contributions of hadronic energy deposited in the 
    electromagnetic section of the MIRAC 
    and the fraction of the electromagnetic
    energy deposited in the lead converter of the PMD. This is estimated to
    vary from 8.8\% to 10.5\% from peripheral to central events.
    Thus the combined systematic error on 
    $\langle p_T\rangle$ has been estimated
    to vary from 11.5\% to 12.7\%.

\section{Results and Discussions}
    The photon multiplicity is determined using equation~(3)
    on an event-by-event basis from
    the total number of $\gamma$-like clusters within the PMD coverage.
    The resulting minimum bias
    distributions of $N_\gamma$
    are shown in Figure~\ref{ngamma} for Pb+Ni, Pb+Nb, and Pb+Pb reactions
    at 158$\cdot A$ GeV. 
    For comparison, the
    corresponding photon multiplicity distributions from the VENUS
    4.12 event 
    generator with default parameter settings, are
    superimposed in the same figure. 
    The shape of these three distributions is governed 
    by the collision geometry.
    For asymmetric collisions of Nb and Ni targets, small shoulders are
    present around $N_\gamma$ of 300 and 200, respectively.
    This shoulder is produced when a decrease in the impact parameter
    leads to little increase in particle production and the cross
    sections for these small impact parameters pile up at a fixed
    $N_\gamma$. The VENUS event generator does not reproduce this
    shoulder well. 
    The $N_\gamma$ values are higher for the data compared to those of
    VENUS.

    The pseudorapidity distribution of photons at different
    centralities are shown in Figure~\ref{eta}~(a) for
    Pb+Pb collisions. 
    The data have been corrected for geometry,
    efficiency, and purity factors. 
    The filled symbols represent
    the measured data, and the open symbols are reflections of the
    filled symbols at $\eta_{c.m.}(=2.95)$.
    The histograms show the corresponding distributions obtained from
    the VENUS event generator. The discrepancy between the 
    VENUS results and the data is about 10\% for central collisions
    at midrapidity.
    The pseudorapidity distribution of photons at different 
    centralities for Pb+Nb and Pb+Ni are shown in Figure~\ref{eta}~(b)
    and (c), respectively.
    The discrepancies between the data and VENUS are larger for these 
    reactions compared to that of Pb+Pb. 

    The pseudorapidity distributions have been fitted with Gaussian
    distributions. The peak position of the distribution ($\eta_{peak}$),
    the pseudorapidity density ($\rho_{max}$), and 
    the width ($\sigma$) are extracted from the fits.
    The $\eta_{peak}$ values for Pb+Pb distributions remain constant
    at 2.95 for all centralities. For Pb+Nb, $\eta_{peak}$
    decreases from $3.03\pm0.15$ 
    for central collisions to $2.95\pm0.32$ in the 
    case of peripheral collisions.
    The corresponding values for the Pb+Ni reaction are $3.10\pm0.16$ 
    and $2.99\pm0.31$. Figure~\ref{etapar}(a) and (b) show
    $\rho_{max}$ and $\sigma$ for the three reactions 
    as functions of the number of participant nucleons, $N_{part}$, at
    different centralities. The $N_{part}$ values
    are determined from the VENUS event generator.
    $\rho_{max}$ increases with $N_{part}$ 
     while $\sigma$ doesn't change with increase of $N_{part}$.

    More insight into the systematics of the particle production 
    can be obtained
    by computing the integrated number of photons ($N_\gamma^{tot}$) 
    over the full phase space. This has been
    obtained from the Gaussian fit parameters    
    to the pseudorapidity distributions.  Figure~\ref{etapar}(c) shows the
    extracted values of $N_\gamma^{tot}$ for Pb+Pb 
    as a function of $N_{part}$.
    The solid line shows a fit to the data using the function:
\begin{equation}
    N_\gamma^{tot} = C \cdot (N_{part})^{\alpha}
\end{equation}
    where $C$ is a proportionality constant. The value of the exponent, 
    $\alpha$, 
    is extracted to be $1.13\pm0.03$.
    To further explore the systematics, we have divided the full $\eta$
    region (0$-$6)
    into two parts, one corresponding to the central rapidity region,
    $2.4\le \eta \le 3.4$, and the other beyond this. 
    For both of these cases, the Pb+Pb data yields a value of 
    $\alpha=1.13$.
    In comparison, fitting the photon distribution from the 
    VENUS event generator in the same two regions yield different 
    exponents, with $1.10\pm0.07$ at mid-rapidity and only 
    $1.0\pm0.05$ for the outer region.

    The mean transverse momentum, $\langle p_T\rangle$, 
    as a function of $\rho_{max}$, the pseudorapidity density of photons 
    at midrapidity, are shown in Figure~\ref{pt} for
    Pb+Pb collisions.
    The data point at $\rho_{max}\simeq525$ corresponds to the highest
    centrality bin, 0$-$1\% of the
    minimum bias cross section for Pb+Pb. The systematic error on the
    absolute values are
    indicated by the upper and lower brackets on the data. The
    $\langle p_T\rangle$ values are constant within the quoted
    error. For comparison, the results obtained
    from the VENUS event generator are superimposed in the figure.
    The $\langle p_T\rangle$ values obtained from VENUS
    are systematically higher compared to data, and 
    show very little change with centrality. The indication of a small
    rise and saturation of
    $\langle p_T\rangle$ seen in the data 
    is similar to what has been reported for
    neutral pions \cite{wa98_pi0}.

\section{Summary}
    Photon multiplicities have been measured, 
    on an event-by-event 
    basis, in the forward region for 158$\cdot A$ GeV Pb induced reactions on 
    Ni, Nb, and Pb targets.
    The peak positions of the photon pseudorapidity distributions 
    are found to 
    shift forward in going
    from the Pb target to Nb and Ni, as expected.
    The photon pseudorapidity densities increase
    with the number of participant nucleons,
    while the widths of the pseudorapidity
    distributions remain constant with centrality. The integrated number of
    photons scales like 
    $N_{part}^{1.13}$, almost independent of the rapidity
    range over which the integration is performed. 
    This is similar to the predictions
    of the VENUS event generator, 
    except that VENUS shows a smaller power for forward rapidities.
    The photon mean transverse momentum has been determined from the 
    event-by-event ratio of the electromagnetic transverse
    energy to the number of photons. 
    After an initial rise with increasing $\rho_{max}$, the value of 
    $\langle p_T\rangle$ is observed to remain constant.
    
\section{Acknowledgements}
We wish to express our gratitude to the CERN accelerator division for
the excellent performance of the SPS accelerator complex. We acknowledge with
appreciation the effort of all engineers, technicians, and support staff who
have participated in the construction of this experiment. 

This work was supported jointly by 
the German BMBF and DFG, 
the U.S. DOE,
the Swedish NFR, 
the Dutch Stichting FOM, 
the Stiftung fuer Deutsch-Polnische Zusammenarbeit,
the Grant Agency of the Czech Republic under contract No. 202/95/0217,
the Department of Atomic Energy,
the Department of Science and Technology,
the Council of Scientific and Industrial Research and 
the University Grants 
Commission of the Government of India, 
the Indo-FRG Exchange Programme,
the PPE division of CERN, 
the Swiss National Fund, 
the INTAS under Contract INTAS-97-0158, 
ORISE, 
Research-in-Aid for Scientific Research
(Specially Promoted Research \& International Scientific Research)
of the Ministry of Education, Science and Culture, 
the University of Tsukuba Special Research Projects, and
the JSPS Research Fellowships for Young Scientists.
ORNL is managed by Lockheed Martin Energy Research Corporation under
contract DE-AC05-96OR22464 with the U.S. Department of Energy.
The MIT group has been supported by the US Dept. of Energy under the
cooperative agreement DE-FC02-94ER40818.

\clearpage

\begin{figure}
\begin{center}
\epsfig{figure=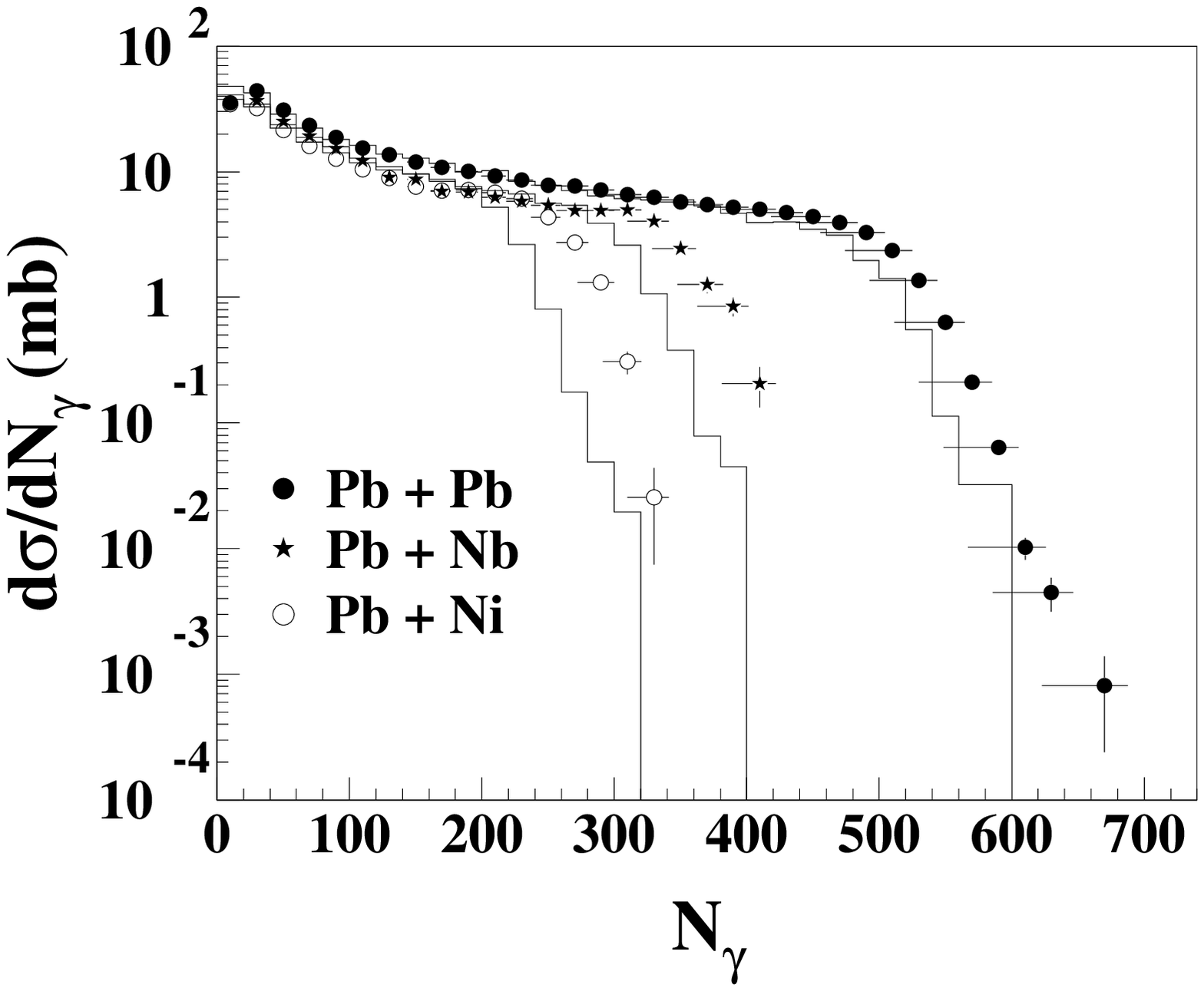,width=16cm}
\caption {\label{ngamma}
Minimum bias inclusive photon cross sections for 
Pb+Ni, Pb+Nb, and Pb+Pb reactions at 158$\cdot A$ GeV.
Solid histograms are the
corresponding distributions obtained from the VENUS event generator.
}
\end{center}
\end{figure}

\clearpage

\newpage
\begin{figure}
\begin{center}
\epsfig{figure=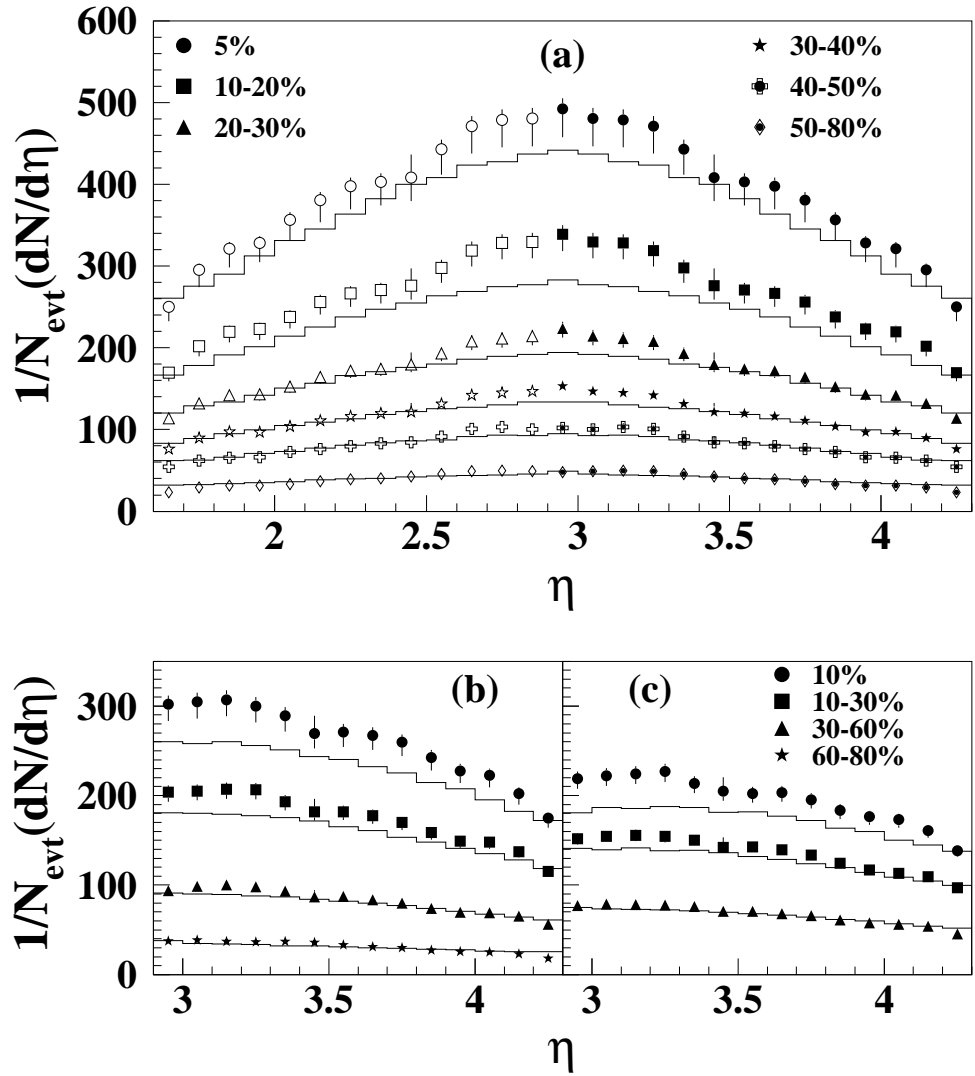,width=14.5cm}
\caption {\label{eta}
Pseudorapidity distributions of photons in Pb induced reactions at
158$\cdot A$ GeV
on (a) Pb, (b) Nb, and (c) Ni targets. The solid histograms
are the corresponding distributions obtained from the VENUS event generator.
}
\end{center}
\end{figure}

\newpage

\begin{figure}
\begin{center}
\epsfig{figure=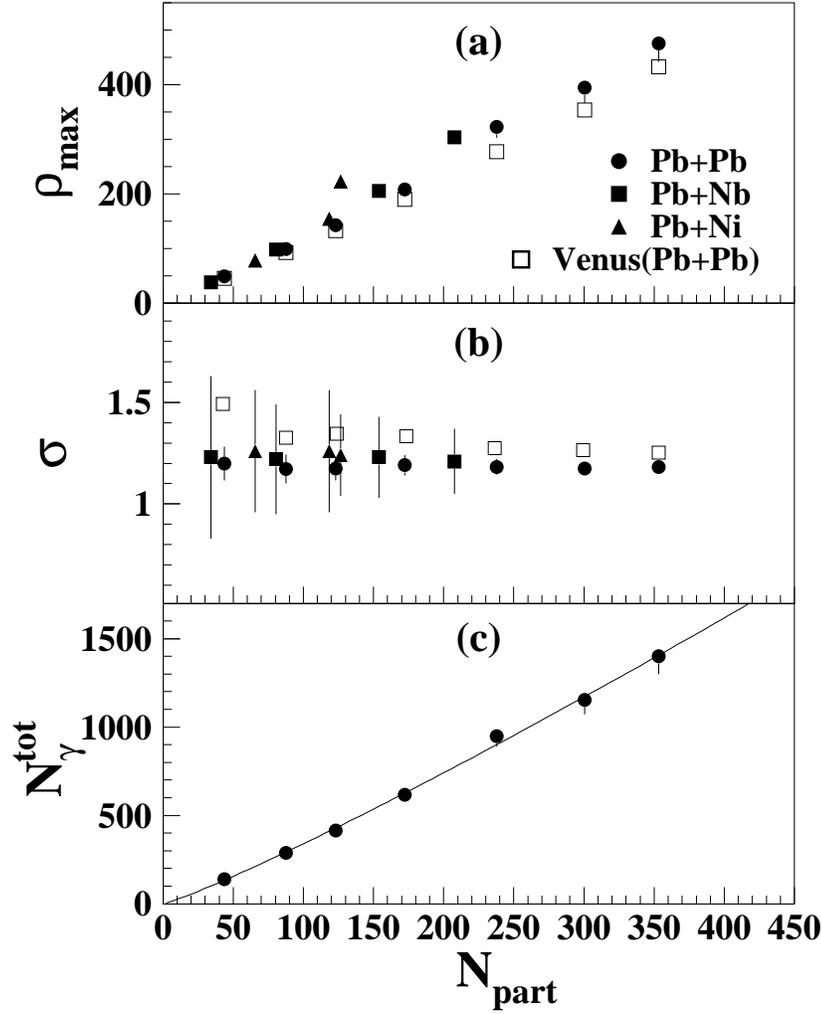,width=12cm}
\caption {\label{etapar}
(a) Pseudorapidity density ($\rho_{max}$), 
(b) width of the pseudorapidity distributions ($\sigma$), and 
(c) integrated values of number of photons ($N_\gamma^{tot}$),
    as functions of the number of participant nucleons at 
    different centrality bins for Pb induced reactions on Ni, Nb, and Pb 
    targets at
    158$\cdot A$ GeV.
    The solid line in (c) is a power-law fit to the data,
    which yields the value of the exponent, $\alpha=1.13\pm0.03$.
}
\end{center}
\end{figure}

\begin{figure}
\begin{center}
\epsfig{figure=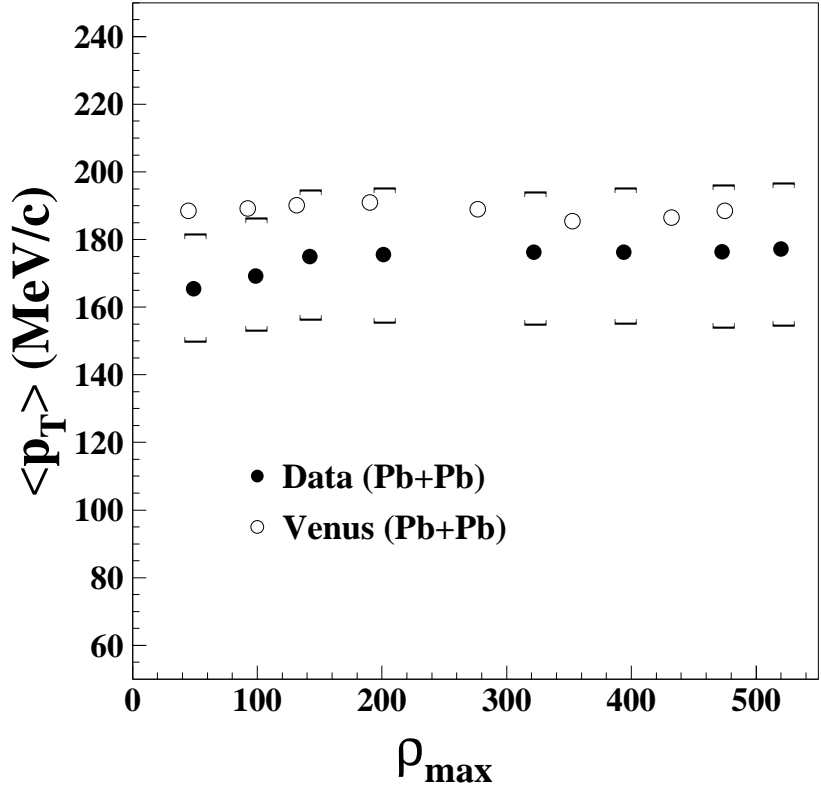,width=12cm}
\caption {\label{pt}
The mean transverse momentum, $\langle p_T\rangle$, of photons as a function of
the pseudorapidity density of photons at midrapidity, $\rho_{max}$,
corresponding to different centralities. The $\langle p_T\rangle$ values
obtained from the 
VENUS event generator for Pb+Pb are superimposed for comparison.
}
\end{center}
\end{figure}

\end{document}